\documentclass[aps,prb,floatfix,superscriptaddress,preprint,showpacs,twocolumn,10pt]{revtex4}

\usepackage[dvips]{graphicx}
\usepackage{graphics}
\usepackage{amssymb}
\usepackage{amsmath}
\usepackage{epsfig}
\usepackage{latexsym}
\usepackage{color}
\usepackage{rotating}

\begin{document}

\title{Continuous quantum non-demolition measurement of Fock states \\ of a nanoresonator using feedback-controlled circuit QED}

\author{M. J. Woolley, A. C. Doherty and G. J. Milburn} 
\affiliation{School of Mathematics and Physics, The University
of Queensland, St Lucia, QLD 4072, Australia}

\begin{abstract}
We propose a scheme for the quantum non-demolition (QND) measurement of Fock states of a nanomechanical resonator via feedback control of a coupled circuit QED system. A Cooper pair box (CPB) is coupled to both the nanoresonator and microwave cavity. The CPB is read-out via homodyne detection on the cavity and feedback control is used to effect a non-dissipative measurement of the CPB. This realizes an indirect QND measurement of the nanoresonator via a second-order coupling of the CPB to the nanoresonator number operator. The phonon number of the Fock state may be determined by integrating the stochastic master equation derived, or by processing of the measurement signal. 
\end{abstract}

\pacs{42.50.Lc,85.85.+j,03.65.Ta}

\maketitle

\section{INTRODUCTION}

An important benchmark for quantum control is the ability to prepare and detect a harmonic oscillator in a Fock state, an energy eigenstate. This may soon be achieved in  nanomechanical resonators \cite{blencowe:QEMS}, which are more usually prepared in quasi-classical displaced thermal states due to the ubiquity of environmental interactions \cite{Schlosshauer}. The ability to prepare a mechanical resonator in a Fock state will provide access to processes that mark the quantum-to-classical transition \cite{martinis:Fockstatedecay}. 

Fock states have already been prepared in other instances of harmonic oscillators. Using a cavity QED system, few-photon Fock states of an electromagnetic mode have been prepared deterministically \cite{walther:cavityQEDFockstates2}, via state reduction \cite{walther:cavityQEDFockstates,haroche:twophotonFockstate} and via quantum non-demolition measurement \cite{guerlin:QNDphotoncounting,brune:cavityQEDFockstates}. Few-photon optical Fock states have also been generated conditionally using parametric down conversion \cite{yamamoto:PDCnumberstates}. Taking advantage of the higher level of control afforded by circuit QED, Fock states containing up to fifteen photons have also been deterministically generated \cite{martinis:Fockstatedecay, schoelkopf:singlephotons, martinis:circuitQEDFockstates}. The motional state of a trapped ion provides another instance of a harmonic oscillator, and indeed Fock states have been prepared by tuning the interaction between the ion's motional state and its internal electronic levels via an external laser \cite{wineland:nonclassicalstatesion}.  

One may consider analogous schemes for the generation of such states of macroscopic mechanical resonators. Indeed, in an extraordinary experiment, Fock states of a mechanical oscillator, in the form of a suspended film bulk acoustic resonator resonantly coupled to a superconducting phase qubit, have been demonstrated \cite{cleland:singlephononcontrol}. However, it would still be desirable to generate mechanical Fock states in a lower-frequency mechanical resonator with a clear separation of mechanical and electromagnetic degrees of freedom. In order to conditionally prepare a Fock state of such a resonator, one requires a coupling to its number operator. In general, this is difficult to realize, though an optomechanical ``membrane-in-the-middle'' system does so \cite{harris:numbercoupling}. In other cases, a key ingredient is the coupling of the resonator to a qubit. Coupling between a nanoresonator and a superconducting qubit has recently been demonstrated by LaHaye \emph{et al.} \cite{lahaye:NRqubit}. Irish and Schwab proposed a scheme for the generation and detection of a Fock state of a nanoresonator via coupling to a Cooper pair box \cite{Irish1}. Jacobs \emph{et al.} have considered the continuous measurement of a nanoresonator using circuit QED \cite{jacobs}; the nanoresonator is coupled to the qubit, and the qubit is read out using a superconducting microwave cavity \cite{schoelkopf}. Here the linear coupling betweeen the nanoresonator and the qubit is brought into resonance by driving the nanoresonator at a frequency equal to the detuning between the nanoresonator and the qubit. It is shown that, in spite of the linear coupling, the effect of the measurement is to localize the nanoresonator onto its Fock states. A dispersive coupling to a qubit with the nanoresonator number operator has also been considered, with the proposed read-out in this case being through a single-electron transistor or a quantum point contact \cite{jacobs:phasestatesNR}. The measurement of Fock states by direct coupling to a single-electron transistor has also been investigated \cite{martin:energyeigenstatesslowdetector}.  Alternative proposals for conditional Fock state generation in nanoresonators are based on engineered cross-Kerr interactions \cite{Santamore1,Buks1}, in analogy with the situation commonly encountered in an optical setting \cite{Yamamoto1}. 

The system under consideration here is composed of a nanoresonator, a superconducting microwave cavity and a Cooper pair box. Both the nanoresonator and the microwave cavity are separately coupled to the Cooper pair box (CPB) \cite{nakamura:CPB}. The direct coupling between the nanoresonator and the cavity could also be considered, and it may be shown that resolved sideband is achievable using a similar analysis \cite{woolley:nanosqueezing}. There is a resonant second-order coupling between the nanoresonator and the CPB, which gives a Hamiltonian coupling between the nanoresonator number operator and the qubit inversion $\sigma_z$ (in the energy eigenbasis of the qubit) as $b^\dagger b \sigma_z$. In the adiabatic limit, the qubit coherence $( \sigma_- )$ may be continuously read-out via homodyne detection on the output of the microwave cavity, and via feedback control of the CPB charge bias a non-dissipative measurement of $\sigma_y$ may be created \cite{wiseman:QNDbyFB}. Such feedback has been considered before in the context of stabilizing pure states of a qubit \cite{wang:stabilizestates}. Thus we have an indirect measurement of the nanoresonator phonon number, and may conditionally prepare a Fock state of the nanoresonator \cite{jacobs}. In principle, the conditional evolution of the system may be reconstructed by integrating the stochastic master equation, though the phonon number may be extracted (and quantum jumps observed) from the oscillations of $\sigma_y$ by filtering the homodyne measurement record. Note that feedback control of nanomechanical systems has previously been considered for the purpose of cooling the mechanical mode toward its ground state \cite{hopkins:FBcoolingOfNanoresonator,ruskov:FBsqueezingOfNanoresonator} and for squeezing of its quantum fluctuations \cite{clerk:squeezingByFeedback,nori:squeezingByIndirectFB}.

\section{SYSTEM AND HAMILTONIAN}

The Hamiltonian describing this system, in the Schr\"{o}dinger picture and in the charge basis of the CPB, is
\begin{eqnarray}
\bar{H}_S & = & \hbar \omega_c a^\dagger a + \hbar \omega_m b^\dagger b + \frac{\hbar \epsilon}{2}\bar{\sigma}_z + \frac{\hbar \Delta}{2}\bar{\sigma}_x \nonumber \\ & & + \hbar  g \left( a + a^\dagger \right) \bar{\sigma}_z + \hbar \lambda \left(b + b^\dagger \right)\bar{\sigma}_z \label{SPHam} 
\end{eqnarray}
where $\omega_c $ is the resonance frequency of the cavity, $\omega_m$ is the resonance frequency of the nanoresonator, and the CPB is described by $\hbar \epsilon = -4E_C(1-2n_g)$ and $\hbar \Delta = -E_J \cos \left(\pi \Phi_e /\Phi_0 \right)$. Here $ E_C = e^2/2C_\Sigma $ is the charging energy of the box, $n_g = n^m_g + n^c_g $, $n^m_g = C^m_g V^m_g /2e$ is the dimensionless gate charge due to the nanoresonator voltage, $n^c_g = C^c_g V^c_g /2e$ is the dimensionless gate charge due to an additional control voltage, $E_J = I_c \Phi_0 /2\pi$ is the Josephson energy of each of two Josephson junctions between the CPB and its superconducting reservoir, $\Phi_e$ is the external flux threaded through the two Josephson junctions, $\Phi_0$ is the magnetic flux quantum and $I_c$ is the critical current of each Josephson junction to the CPB. The total capacitance of the box is given by $C_\Sigma = 2C_J + C_g + C^m_g + C^c_g$ where $C_J$ is the capacitance of each Josephson junction, $C_g$ is the coupling between the box and the central conductor of the microwave cavity, $C^m_g$ is the capacitance between the nanoresonator and the box, and $C^c_g$ is the capacitance between the control gate and the box. 

It is assumed that the CPB is configured as a charge qubit, meaning that $E_C >> E_J $. The couplings between the CPB and the cavity \cite{Blais1}, and between the CPB and the nanoresonator \cite{Armour} are given by, respectively,
\begin{equation}
\hbar g = e \frac{C_g}{C_\Sigma} \sqrt{ \frac{\hbar \omega_c}{ cL } }, \ \ \ \hbar \lambda = 4E_C n^m_g \frac{\Delta x}{d} .
\end{equation}
Here $c$ is the capacitance per unit length of the microwave cavity, $L$ is the length of the microwave cavity, $\Delta x$ is the half-width of the nanoresonator ground-state wavefunction, and $d$ is the equilibrium separation of the nanoresonator and the CPB. 

First we rotate into the energy eigenbasis of the uncoupled qubit; the natural basis for a treatment of dissipation. Additionally transforming into an interaction picture with respect to $H_1 = \hbar \omega_c a^\dagger a $, 
\begin{eqnarray}
H_{I_1} & = & \hbar \omega_m b^\dagger b + \frac{\hbar \Omega}{2}\sigma_z + \hbar\lambda \left( b + b^\dagger \right) \left(\frac{\epsilon}{\Omega}\sigma_z - \frac{\Delta}{\Omega}\sigma_x  \right) \nonumber \\ & & \ \ + \hbar  g \left( a e^{-i\omega_c t} + a^\dagger e^{i\omega_c t} \right) \left(\frac{\epsilon}{\Omega}\sigma_z - \frac{\Delta}{\Omega}\sigma_x  \right) , 
\end{eqnarray}
where the CPB energy splitting is $\Omega = \sqrt{\epsilon^2 + \Delta^2 } $. 

In the absence of resonant terms in the coupling between the nanoresonator and the CPB, we make a Schrieffer-Wolff transformation \cite{Hauss1} to extract a resonant second-order coupling; $H^{SW}_{I_1} = S^\dagger H_{I_1} S$ where $S = {\rm exp} \left[ i (\lambda \Delta /\Omega^2) \sigma_y \left( b + b^\dagger \right) \right]$. This is essentially equivalent to considering the dispersive limit of the coupling between a qubit and an oscillator. Retaining terms to linear order in $\lambda \Delta / \Omega^2 $, transforming to another interaction picture with respect to $H_2 = \hbar \left( \omega_c/2 \right) \sigma_z + \hbar \omega_m b^\dagger b $, and employing the rotating-wave approximation to neglect rapidly oscillating terms, we find 
\begin{eqnarray}
H_{I_2} & = & \hbar \left( \delta + \chi b^\dagger b \right) \sigma_z + \hbar g' \left(a\sigma_+ + a^\dagger \sigma_- \right) \label{firstHam} ,
\end{eqnarray}
where $\delta = \left( \Omega - \omega_c \right)/2 $ is the detuning between the CPB and the cavity, $g' = -g\Delta / \Omega$ is the effective coupling to the cavity, and $\chi = 4\lambda^2\Delta^2 / \Omega^3 $ is the effective coupling to the nanoresonator. The cavity has a Jaynes-Cummings coupling to the CPB, which itself is dispersively coupled to the nanoresonator. 

\section{CONDITIONAL DYNAMICS UNDER HOMODYNE DETECTION}

We consider the conditional evolution of the system described by Eq.~(\ref{firstHam}) under homodyne detection of the field output from the microwave cavity. There exist proposals for near quantum-limited measurement of the quadratures of a microwave field \cite{sarovar,teufel}. It is assumed that the microwave field, leaking out of the cavity at a rate $\mu$, is subject to homodyne detection with respect to a local oscillator phase $\theta$ and with an efficiency $\eta$, that the nanoresonator is damped at a rate $\gamma$ into a thermal environment characterised by the occupation $n^0_m$, and that the CPB experiences spontaneous emission at a rate $\Gamma_q$. The effect of the Schrieffer-Wolff transformation on dissipative terms may be neglected under the joint assumptions of weak damping and $\lambda \Delta / \Omega^2 << 1$. The master equation describing the conditional evolution of the system is  
\begin{eqnarray}
d\rho & = & -\frac{i}{\hbar}\left[ H_{I_2} , \rho \right]dt + \gamma \left( n^0_m + 1 \right)\mathcal{D}\left[ b \right] \rho dt \nonumber \\ & & \ \ \ \  + \gamma n^0_m \mathcal{D}\left[ b^\dagger \right] \rho dt + \mu \mathcal{D}\left[ a \right]\rho dt + \Gamma_q \mathcal{D}\left[ \sigma_- \right]\rho dt \nonumber \\ & & \ \ \ \ \ \ +    \sqrt{\eta \mu} \ \mathcal{H} \left[ ae^{-i\theta} \right] \rho \ dW  ,  
\end{eqnarray}
with $dr = \left\langle ae^{-i\theta} + a^\dagger e^{i\theta} \right\rangle dt + dW / \sqrt{\eta \mu}$ as the measurement record increment (where $dW$ is the Wiener increment), $ \mathcal{D}(s)\rho = s\rho s^\dagger - \frac{1}{2}s^\dagger s \rho - \frac{1}{2}\rho s^\dagger s $ as the dissipative superoperator and $ \mathcal{H}\left[ s \right]\rho = s\rho + \rho s^\dagger - {\rm tr}\left(s\rho + \rho s^\dagger \right)\rho $ as the measurement superoperator \cite{wiseman:quadraturemeasurement}. 

In the good-measurement limit, the cavity damping rate is large, $\mu >> \delta , \chi , g' $, and we may adiabatically eliminate the cavity mode. This is easily done by solving the quantum Langevin equation in the limit of large damping to find $a = \left( -2ig'/\mu \right)\sigma_- $ . Substitution into the conditional master equation above leads to a conditional master equation for the reduced density operator describing the CPB and the nanoresonator, $\rho_r (t) = {\rm tr_{cavity}} \ \rho (t)$, 
\begin{eqnarray}
d\rho_r & = & -i\left[ \left( \delta + \chi b^\dagger b \right) \sigma_z , \rho_r \right] \ dt + \Gamma_q \mathcal{D}\left[\sigma_- \right] \rho_r \ dt \nonumber \\ & &+ \gamma \left( n^0_m + 1 \right) \mathcal{D}\left[ b \right]\rho_r \ dt + \gamma n^0_m \mathcal{D}\left[ b^\dagger \right] \rho_r \ dt \nonumber \\ & &+ \Gamma \mathcal{D} \left[ \sigma_- \right] \rho_r dt + \sqrt{\eta \Gamma} \mathcal{H} \left[ \sigma_- e^{ -i ( \theta + \pi / 2 ) } \right] \rho_r dW . \label{eq:redME}
\end{eqnarray}
where the measurement rate is $\Gamma = 4g'^2/\mu$, and the measurement record increment (setting $\theta = - \pi $) is
\begin{equation}
dr = \sqrt{ \frac{\Gamma}{\mu} } \left\langle \sigma_y \right\rangle dt + \frac{dW}{\sqrt{\eta \mu}} . 
\end{equation} 
The associated measurement current is defined through $dr \equiv I(t) \ dt$.

\section{FEEDBACK CONTROL}

Now Eq.~(\ref{eq:redME}) describes a dissipative measurement of the CPB; the measurement perturbs the observable from which we wish to extract information about the nanoresonator. One may consider using state-based feedback control \cite{doherty:statebasedFB} of the CPB to optimally extract information from the measurement; for example, one may lock the net detuning $(\delta + \chi b^\dagger b )$ to zero such that the CPB observable with a reactive response to the detuning exhibits optimal sensitivity to the nanoresonator state. 

However, a simpler alternative is afforded by Hamiltonian feedback control using the measurement record \cite{wiseman:homodyneFB}. Assuming that the feedback may be applied fast on the time-scale of system dynamics, we may apply the Markovian homodyne feedback master equation. Consider a density matrix $\sigma$ evolving according to the homodyne measurement master equation in Lindblad form,
\begin{equation} 
d\sigma = - \frac{i}{\hbar} \left[ H, \sigma \right] dt + k \mathcal{D} \left[ c \right] \sigma dt + \sqrt{\eta k} \mathcal{H} \left[ c \right] \sigma \ dW ,
\end{equation}
and having the associated homodyne measurement current,
\begin{equation}
I(t) = \left\langle c + c^\dagger \right\rangle (t) + \xi (t) /\sqrt{\eta k} ,
\end{equation}
where $k$ is the measurement rate and $\xi (t) = dW/dt$.
Then, if one applies Markovian feedback using the homodyne measurement current according to the Hamiltonian,
\begin{equation}
H_{fb}(t) = \hbar \lambda F I(t) , \label{eq:FBHamiltonian}
\end{equation}
then according to the quantum theory of continuous feedback \cite{wiseman:quantumFB}, the system evolution is given by
\begin{eqnarray}
d\sigma & = &  - \frac{i}{\hbar} \left[ H, \sigma \right] dt - \frac{i}{2} \lambda \left[ c^\dagger F + F c , \sigma \right] dt \nonumber \\
& & + \mathcal{D} \left[ \sqrt{k} c - i \frac{\lambda}{\sqrt{k}} F \right] \sigma dt + \frac{\lambda^2}{k} \frac{1-\eta}{\eta} \mathcal{D} \left[ F \right] \sigma dt  \nonumber \\ 
& & + \mathcal{H} \left[ \sqrt{\eta k} c - i \lambda F/\sqrt{\eta k } \right] \sigma dW.  \label{eq:feedbackME}
\end{eqnarray}

Now applying the Hamiltonian feedback of Eq.~(\ref{eq:FBHamiltonian}) to the system described by Eq.~(\ref{eq:redME}) with the identification $c \equiv i \sigma_-$ in Eq.~(\ref{eq:feedbackME}), we find
\begin{eqnarray}
d\rho_r & = & -i\left[ \left( \delta + \chi b^\dagger b \right) \sigma_z , \rho_r \right] \ dt + \Gamma_q \mathcal{D}\left[\sigma_- \right] \rho_r \ dt \nonumber \\ 
& & + \gamma \left( n^0_m + 1 \right) \mathcal{D}\left[ b \right]\rho_r \ dt + \gamma n^0_m \mathcal{D}\left[ b^\dagger \right] \rho_r \ dt \nonumber \\ 
& & + \frac{\lambda^2}{\Gamma} \frac{1-\eta}{\eta} \mathcal{D} \left[ F \right] \rho_r dt + \mathcal{D} \left[ \sqrt{\Gamma} \sigma_- - \frac{\lambda}{\sqrt{\Gamma}} F \right] \rho_r dt \nonumber \\
& & + \frac{\lambda}{2} \left[ F \sigma_- - \sigma_+ F, \rho_r \right] dt \nonumber \\ 
& & + \mathcal{H} \left[ i \sqrt{\eta \Gamma} \sigma_- - i\lambda F/\sqrt{\eta \Gamma} \right] \rho_r dW . \label{eq:redMEwithFB}
\end{eqnarray}
In general, the feedback will alter the Hamiltonian, dissipative and measurement dynamics. 


\section{Non-Dissipative Measurement}
Now it is possible, in the case of unit detection efficiency, to create a non-dissipative measurement of $\sigma_y$ from a dissipative measurement \cite{wiseman:QNDbyFB}. In order to achieve such a measurement here, we require that $\lambda F = (\eta \Gamma /2) \sigma_x$. The corresponding feedback Hamiltonian, given by Eq.~(\ref{eq:FBHamiltonian}), is in a rotated qubit basis and an interaction picture, as per Eq.~(\ref{firstHam}). In the Schr\"{o}dinger picture and charge basis of Eq.~(\ref{SPHam}), the required feedback term is $-\hbar \mathcal{E}_{fb}(t) \cos \Omega t \bar{\sigma}_z$ where $\mathcal{E}_{fb}(t) = \left( \eta \Gamma \Omega / \Delta \right) I(t)$. That is, the homodyne measurement record must be used to modulate the CPB charge bias, controlled through $V^c_g$, to create a non-dissipative measurement of $\sigma_y$. 

The associated homodyne feedback master equation is 
\begin{eqnarray}
d\rho_r & = & -i\left[ \left( \delta + \chi b^\dagger b \right) \sigma_z , \rho_r \right] \ dt + \Gamma_q \mathcal{D}\left[\sigma_- \right] \rho_r \ dt \nonumber \\ & & \ \ + \gamma \left( n^0_m + 1 \right) \mathcal{D}\left[ b \right]\rho_r \ dt + \gamma n^0_m \mathcal{D}\left[ b^\dagger \right] \rho_r \ dt \nonumber \\ & & \ \ \ \ + \Gamma \mathcal{D} \left[ \sigma_- - \frac{\eta}{2} \sigma_x \right] \rho_r dt + \left( 1-\eta \right) \frac{\Gamma \eta}{4} \mathcal{D}\left[ \sigma_x \right] \rho_r dt  \nonumber \\ & & \ \ \ \ \ \ + \sqrt{\eta \Gamma} \mathcal{H} \left[ \sigma_y / 2 \right] \rho_r \ dW . \label{eq:homodyneFBME}
\end{eqnarray}
Here there is no alteration to the Hamiltonian dynamics, though for inefficient detection, the feedback also introduces a dephasing of $\sigma_x$, and the measurement becomes dissipative. 

\begin{figure}[tbh]
\includegraphics[scale=0.2]{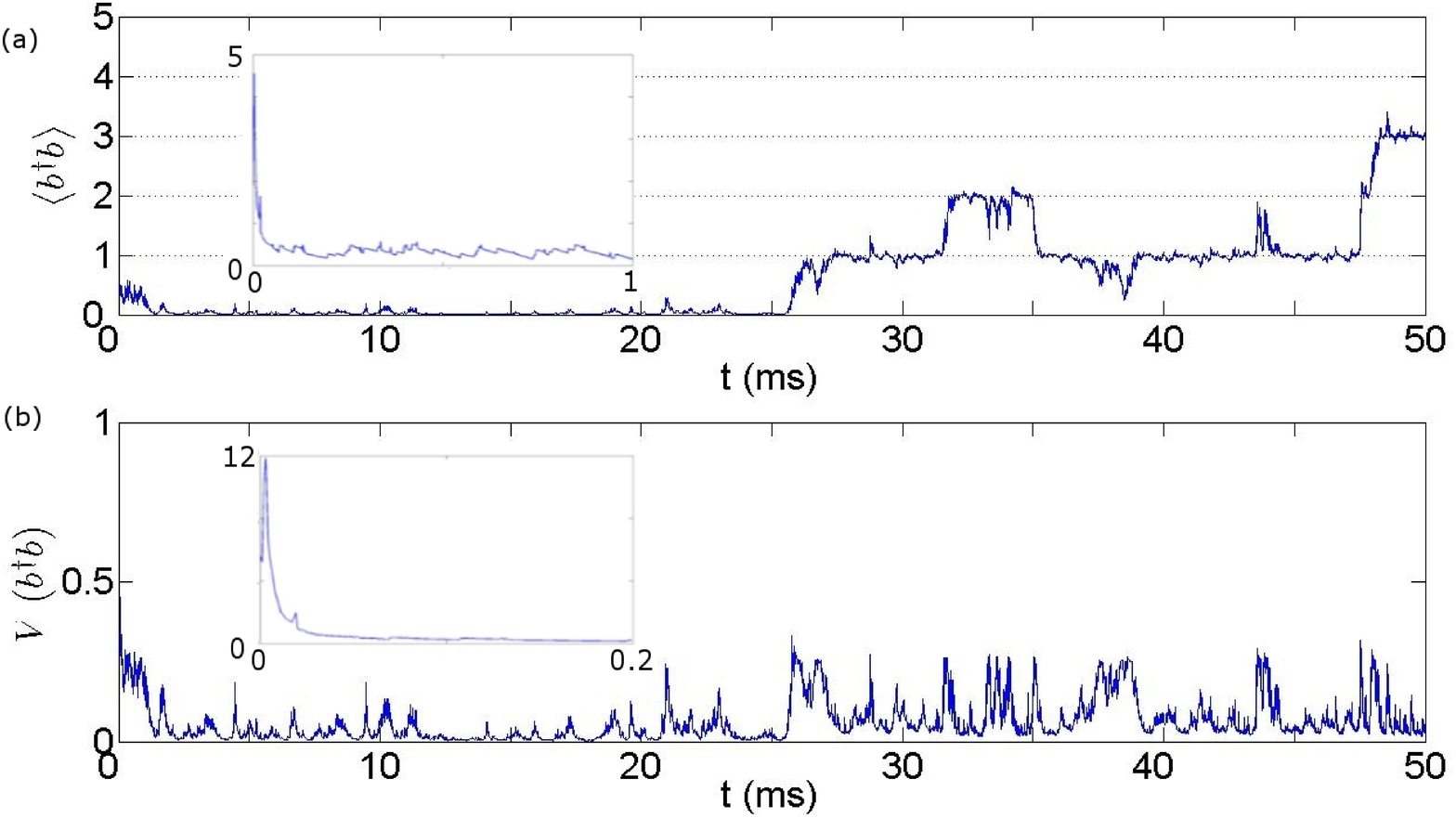}
\caption{Evolution of: (a) phonon number $\left\langle b^\dagger b \right\rangle $ and (b) phonon number variance ${\rm Var} \left( b^\dagger b \right) \equiv \langle \left( b^\dagger b \right)^2 \rangle - \left\langle b^\dagger b \right\rangle^2 $ for a typical trajectory assuming unit detection efficiency $\left( \eta = 1 \right)$. Parameters are as given in the body of the text. The inset shows the conditioning onto a Fock state that occurs on a time-scale short compared to that required for the observation of retroactive quantum jumps. }
\end{figure}

\section{SIMULATIONS}

It is assumed that the nanoresonator has a resonance frequency $\omega_m /2\pi = 10{\rm MHz}$ and a mass $m = 10^{-15}{\rm kg}$, corresponding to ground-state half-width of $\Delta x = 29 {\rm fm} \ $. These are reasonable parameters for a nanoresonator fabricated within a superconducting microwave cavity \cite{schwab:nearground}. The CPB is characterized by $\Delta = 5 \times 10^{10} {\rm s^{-1}}$, $n^m_g = 1/2$ such that $\epsilon = 0$, and hence $\Omega = 5 \times 10^{10}{\rm s^{-1}}$, resonant with the cavity at $\omega_c = 5\times 10^{10}{\rm s^{-1}}$; such values are typical for circuit QED experiments \cite{schoelkopf}. The cavity damping rate is $\mu = 10^7 {\rm s^{-1}}$, and the assumed mechanical quality factor is, an admittedly high, $2\times 10^7$. The bare qubit-nanoresonator coupling is $\lambda /2\pi = 1.5{\rm MHz} $, and the bare qubit-cavity coupling is $g/2\pi = 200{\rm kHz} $, both of which are achievable \cite{lahaye:NRqubit,schoelkopf}. The effective CPB-nanoresonator coupling is $ \chi = 2.56 \times 10^3 {\rm s^{-1}} $, and the effective qubit-cavity coupling is $ g' = -7.56 \times 10^5 {\rm s^{-1}}$. The effect of the cavity is quantified by the measurement rate $ \Gamma = 2.29 \times 10^5 \rm{s^{-1}} $, and spontaneous emission from the qubit occurs at a rate $\Gamma_q = 10^4{\rm s^{-1}}$.  

Simulations of the mixed-state evolution of the nanoresonator-qubit system according to Eq.~(\ref{eq:homodyneFBME}) were performed using an explicit Milstein scheme \cite{kloeden:numericalSDE}. Since this is an instance of a partially-observed system, conventional pure-state unravellings \cite{carmichael} of the master equation are not applicable, though techniques for pure-state unravellings using the concept of ostensible trajectories have been developed \cite{wisemangambetta}. 

\begin{figure}[tbh]
\includegraphics[scale=0.195]{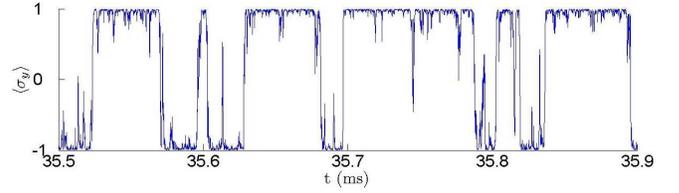}
\caption{The interaction picture evolution of the qubit observable $\left\langle \sigma_y \right\rangle$ corresponding to a short interval of the simulation shown in Fig.~1. The dynamics are the result of a compromise between projection onto the eigenstates of $\sigma_y$ and oscillation due to the free Hamitonian, determined by the relative magnitude of $\Gamma$ and $\chi \left\langle b^\dagger b \right\rangle$. }
\end{figure}

Fig.~1(a) and Fig.~1(b) show, respectively, the evolution of $\left\langle b^\dagger b \right\rangle$ and ${\rm Var} \left( b^\dagger b \right)$ over the course of a typical quantum trajectory assuming unit detection efficiency. The initial state in this case is a thermal state with $\bar{n}=2$ and no coherent amplitude. This is close to the level of cooling recently achieved in a microwave cavity optomechanical system through resolved sideband cooling \cite{schwab:nearground}. From Fig.~1(a) it is seen that the effect of the measurement is to condition the nanoresonator onto a state with an integer $\left\langle b^\dagger b \right\rangle$, and from Fig.~1(b) it is seen that ${\rm Var} \left( b^\dagger b \right)$ goes to zero, verifying that the nanoresonator has been projected onto a Fock state by virtue of the indirect measurement. In the absence of the measurement, the variance would tend to $n^0_m \left( n^0_m + 1 \right)$, in equilibrium with the thermal mechanical heat bath. After some time, jumps between adjacent Fock states are observed, corresponding to the emission and absorption of quanta to/from the thermal mechanical heat bath. These have previously been termed retroactive quantum jumps \cite{wiseman:retroactivejumps}, and arise from the interplay of system and measurement dynamics, in spite of the lack of explicit jump terms in the stochastic master equation. 

Fig.~2 shows the interaction picture evolution of the measured observable $\sigma_y$ over a short interval of the simulation shown in Fig.~1. The measurement acts to put the qubit into an eigenstate of $\sigma_y$, while the free Hamiltonian $\sigma_z$ component drives oscillations of $\left\langle \sigma_y \right\rangle$ at a frequency dependent on the phonon number of the nanoresonator. The dominant effect is determined by the relative magnitudes of the measurement rate $\Gamma$ and the Hamiltonian coupling coefficient $\chi \left\langle b^\dagger b \right\rangle$. In Fig.~2, both effects are observable, and it is clear that by monitoring $\left\langle \sigma_y \right\rangle$, one acquires information about $\left\langle b^\dagger b \right\rangle$. 

\begin{figure}[tbh]
\includegraphics[scale=0.195]{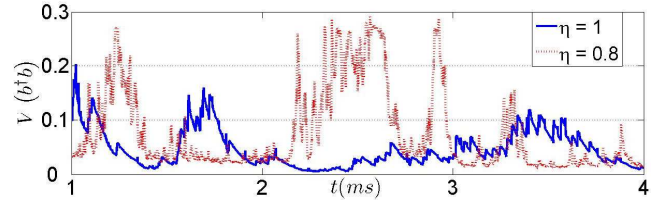}
\caption{Comparison of phonon number variance for typical trajectories following projection onto a Fock state, for efficient and inefficient detection. Greater fluctuations are observed in the case of inefficient detection. }
\end{figure} 

In the presence of detector inefficiency, one still observes projection onto the nanoresonator's Fock states, though a longer time is required for this to be achieved. As shown in Fig.~3, the fluctuations in phonon number variance observed in the case of inefficient detection tend to be greater than those observed in the case of efficient detection. 

\section{CONCLUSIONS}

Microwave cavities provide a natural environment for the application of quantum control protocols; the emission of a qubit can be effectively confined by the microwave cavity, and one can feed back onto the qubit or the microwave cavity on a time-scale comparable to the system frequencies. Here we propose a system that exploits this possibility by using feedback control of a Cooper pair box qubit to implement a quantum non-demolition of the Fock state of a nanomechanical resonator. The most significant experimental challenges are the fabrication of mechanical resonators with sufficiently high quality factors, and the efficient measurement of a quadrature of the microwave cavity field.

\end{document}